\documentstyle[11pt,moriond_preprint,epsfig]{article}

\begin{document}
\begin{flushright}
IPHE 2000-009 \\ May 18, 2000
\end{flushright}
\vspace*{4cm}
\title{HEAVY FLAVOUR PHYSICS RESULTS FROM LEP 1}

\author{~ \\ O.\ SCHNEIDER}

\address{~ \\ Institut de Physique des Hautes Energies, \\
Universit\'e de Lausanne, 
CH-1015 Lausanne, Switzerland ~\\ 
\mbox{\footnotesize e-mail: Olivier.Schneider@iphe.unil.ch} } 

\maketitle\abstracts{
Recent heavy flavour results from the LEP experiments are presented. 
These include a search for new physics in rare $B$ decays, 
a new model-independent measurement of the $b$-quark fragmentation function
at the $Z$ peak, updated measurements of $|V_{cb}|$, 
results on $\Delta \Gamma_s$, searches for $B_s^0$ 
oscillations, as well as a new measurement of $\sin(2\beta)$. 
Many combined results, obtained by dedicated working groups, 
are also given.
The LEP measurements of $V_{cb}$ from $B^0 \to D^{*-}\ell^+\nu$ decays 
average to $|V_{cb}| = (39.7 \pm 3.0)\times 10^{-3}$, while inclusive 
measurements yield $|V_{cb}| = (40.8 \pm 2.0)\times 10^{-3}$ 
dominated by theoretical uncertainties. Charmless semileptonic decays
have been observed inclusively, 
${\rm BR}(b \to u\ell^-\bar{\nu}) = (1.67 \pm 0.55)\times 10^{-3}$, 
corresponding to 
$|V_{ub}| = (4.04 ^{+0.62}_{-0.74}) \times 10^{-3}$.
Significant progress has been made in the $B_s^0$ sector, where the width 
difference is now close to being measured with a combined result of 
$\Delta\Gamma_s/\Gamma_s = 0.24 ^{+0.16}_{-0.12}$ or 
$\Delta\Gamma_s/\Gamma_s < 0.53$ at 95\%~CL.
However, despite continuing improvements, $\Delta m_s$ is still unmeasured, 
with a lower limit of 14.6~ps$^{-1}$ at  95\%~CL.
Improved heavy flavour results are expected from LEP by Summer 2000.
The current status of electroweak heavy flavour physics is summarized in 
another presentation.}

\vfill
\begin{center}
{\em Talk presented at the XXXVth Rencontres de Moriond,} \\ 
{\em Electroweak Interactions and Unified Theories,} \\
{\em Les Arcs, France, March 11--18, 2000.} \\ ~
\end{center}
\newpage

It is not possible to review here all the heavy flavour results produced 
by the ALEPH, DELPHI, L3 and OPAL experiments based on their LEP 1 data, 
taken until 1995 at an $e^+e^-$ center-of-mass energy equal or close to 
the $Z$ mass. Although the peak activity of LEP 1 analysis is behind us, 
these data are still being analyzed, producing new and improved 
measurements. 

The focus of this presentation is on new or updated 
$b$-physics results released since the 1999 Summer Conferences, as well as
on the latest averages produced by various LEP heavy flavour working 
groups.\,\cite{LEPHF}
Many of these results can be related to the magnitude of the least well 
known CKM matrix elements $|V_{cb}|$, $|V_{ub}|$, $|V_{td}|$ and $|V_{ts}|$, 
which are in turn related to the lengths of the sides of the 
CKM unitarity triangle. 
These measurements will be (together with the forthcoming results 
from the $B$ factories and the Tevatron) important ingredients of future tests 
of the CKM picture within the Standard Model, 
where an inconsistency may indeed 
be an indirect indication of new physics. 

New physics may also be responsible for unexpectedly high branching ratios
in rare $B$ decays involving flavour-changing neutral currents.
For example the branching ratio of the decay $B\to K^- K^- \pi^+$,
for which the OPAL collaboration has recently 
released\,\cite{OPAL-search} an upper limit of $1.29 \times 10^{-4}$ 
at 90\% CL, is predicted to be $10^{-11}$ at most in the Standard Model 
(box diagram) but is practically unconstrained in certain supersymmetric 
models with R-parity violation (tree diagram possible).

\section{\boldmath $b$-fragmentation studies}
\label{sec:bfrag}

Understanding the production of $b$-hadrons in $Z$ 
decays is important for many heavy flavour analyses. 
The $b$-quark hadronisation 
can be described in terms of the variable 
$x_E =E_{\mbox{\scriptsize $b$-hadron}}/E_{\rm beam}$, 
the fraction of the beam energy retained by the weakly-decaying 
$b$-hadron produced in a $b$ jet. Being not known accurately, 
the distribution of $x_E$ is simulated by the LEP experiments
using the JETSET generator together with phenomenological models that 
relate the energy of the $b$-hadron with that of the initial $b$-quark. 
The most commonly used of these models, from Peterson et al., 
relies on a single parameter which has merely been tuned to reproduce the 
experimental spectra of high transverse momentum leptons originating 
mostly from the decay $b \to c \ell^- \bar{\nu}$. This tuning 
corresponds to a mean $x_E$ 
of $\langle x_E \rangle = 0.702 \pm 0.008$, 
which is the value recommended up to now
for heavy flavour analyses at LEP.\,\cite{LEPHFWG_1998}

\begin{figure}
\begin{minipage}{0.49\textwidth}
\begin{center}
\epsfig{figure=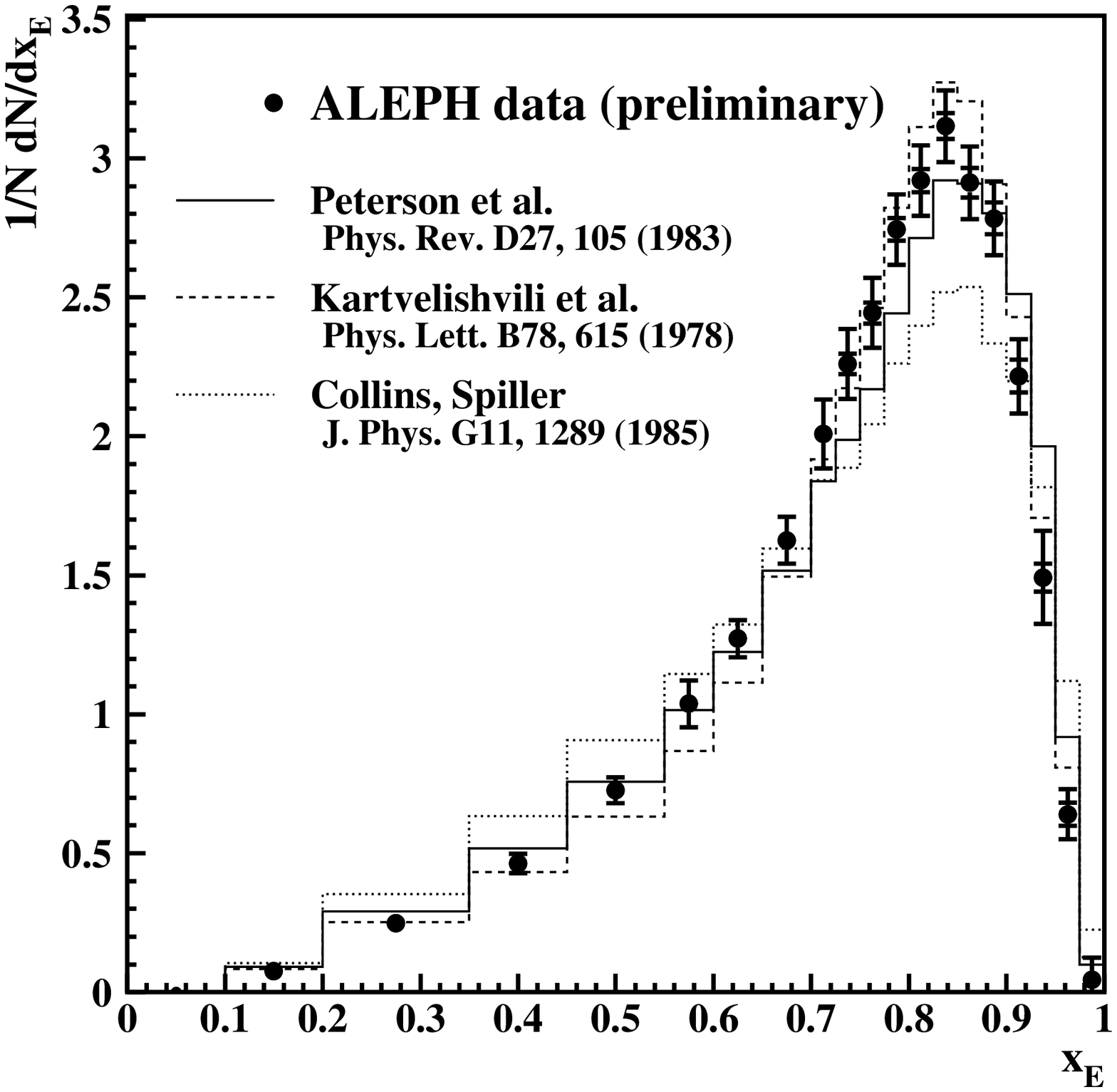,width=1.0\textwidth}
\caption{$x_E$ distribution of $B \to D^{*} l \nu$ mesons
measured by ALEPH\,\protect\cite{ALEPH-bfrag} 
compared with predictions from various $b$-fragmentation models 
fitted to the data.} 
\label{fig:bfrag}
\end{center}
\end{minipage}
\hfill 
\begin{minipage}{0.49\textwidth}
\begin{center}
\epsfig{figure=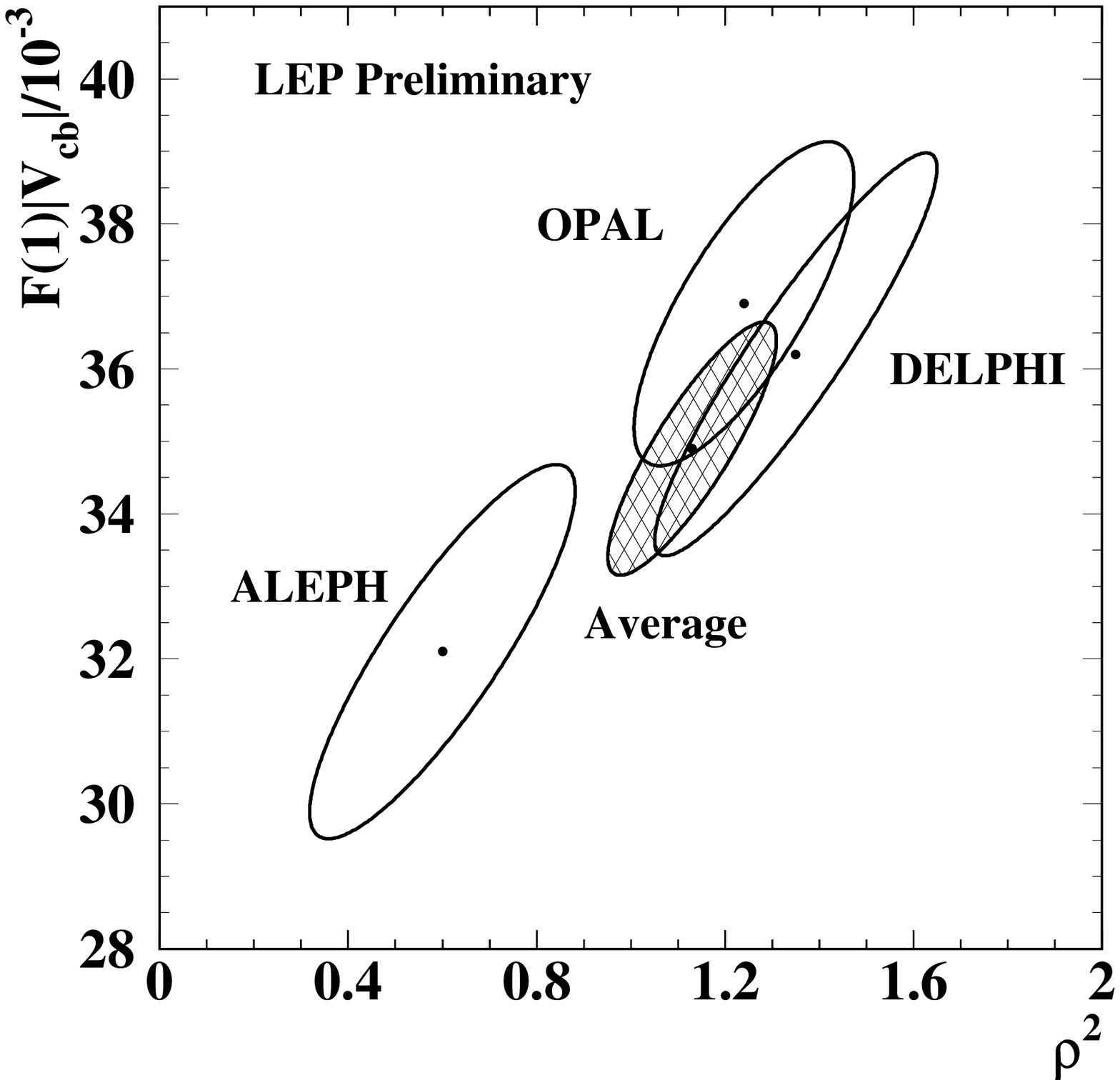,width=1.065\textwidth} \\[-5mm]
\caption{LEP measurements of ${\cal F}_{D^*}(1) |V_{cb}|$ and $\rho^2$ from 
$B^0 \to D^{*-} \ell^+ \nu$ decays, and their average.
The error ellipses (39\% CL) include systematic uncertainties.}
\label{fig:vcb}
\end{center}
\end{minipage}
\end{figure}

The ALEPH collaboration has submitted to this conference 
a new measurement\,\cite{ALEPH-bfrag} of the shape of the $x_E$ 
distribution (see Fig.~\ref{fig:bfrag}) 
based on approximately 3000 $B \to D^{*} l \nu$ decays, 
where the $B$ meson energy has been estimated in a model-independent way 
from an identified lepton, a fully reconstructed 
$D^{*}$ meson and missing energy information. The energy spectrum 
is found to be somewhat harder than assumed before, with 
$\langle x_E \rangle = 0.7198 \pm 0.0045 _{\rm stat} \pm 0.0053 _{\rm syst}$,
consistent with a previous ALEPH measurement. This 
confirms a recent result from SLD,\,\cite{SLD-bfrag} 
$\langle x_E \rangle = 
0.714 \pm 0.005 _{\rm stat} \pm 0.007 _{\rm syst} \pm 0.002 _{\rm model}$,
which has small model-dependent systematics, although based on an inclusive 
sample of $b$-hadrons. 

These direct measurements of the shape of the $x_E$ distribution
have now sufficient precision to envisage tests of the $b$-fragmentation 
model predictions and to discriminate amongst these models for the first 
time. For example, both ALEPH and SLD data favour the description of 
Kartvelishvili et al.\ over the one from Peterson et al.

\section{Measurements of \boldmath $|V_{cb}|$ and $|V_{ub}|$}
\label{sec:vcb-vub}

The study of the $B^0 \to D^{*-} \ell^+ \nu$ decay kinematics allows the
extraction of $|V_{cb}|$. The differential decay rate as a function of the 
boost $\omega$ of the $D^{*-}$ in the $B^0$ rest frame is predicted by the
Heavy Quark Effective Theory to be
\begin{equation}
\frac{d\Gamma}{d\omega} = {\cal K}(\omega) {\cal F}_{D^*}^2(\omega) 
|V_{cb}|^2 \,,
\end{equation} 
where ${\cal K}(\omega)$ is a known phase-space function and 
${\cal F}_{D^*}(\omega)$ a single form factor which, 
in the heavy quark limit, is equal to unity at zero recoil. 
The interesting observable is thus the decay rate at 
zero recoil, but since the phase space vanishes at $\omega=1$, the quantity
${\cal F}_{D^*}(1) |V_{cb}|$ must be extracted from an extrapolation of the 
measured differential rate at $\omega>1$. At LEP, this extrapolation relies 
on a specific parametrization\,\cite{Vcb-param} of the shape of 
${\cal F}_{D^*}(\omega)$ in terms of the slope $\rho^2$ at $\omega =1$.

The OPAL collaboration has recently updated their result obtained with 
fully reconstructed decays and performed a new analysis\,\cite{Vcb-OPAL-new}
based on an inclusive $D^{*-}$ reconstruction relying on the identification 
of the slow pion from the $D^{*-}$ decay. Their new combined result is 
displayed on Fig.~\ref{fig:vcb}, together with earlier and similar
measurements from ALEPH and DELPHI. A combination of these results, 
performed by the LEP $V_{cb}$ working group,\,\cite{LEPHF} 
takes into account all correlations and yields 
\begin{equation}
{\cal F}_{D^*}(1) |V_{cb}| = 
(34.9 \pm 0.7 \pm 1.6) \times 10^{-3} ~~~ \mbox{and} ~~~
\rho^2 = 1.13 \pm 0.08 \pm 0.16 \,.
\end{equation}
Systematic uncertainties are dominated by the limited knowledge of the 
$D^*$ recoil spectrum for $B \to D^{*} \ell \nu X$ background events.
With ${\cal F}_{D^*}(1)=0.88\pm 0.05$ from theoretical calculations taking 
into account finite quark masses and QCD corrections,\,\cite{HF_preprint}
this leads to the combined LEP estimate $|V_{cb}| = (39.7 \pm 0.8 _{\rm stat} 
\pm 1.8 _{\rm syst} \pm 2.2 _{\rm theory}) \times 10^{-3}$ 
from exclusive decays.

Current theoretical calculations based on heavy quark symmetry 
relate $|V_{cb}|$ and $|V_{ub}|$ to the inclusive $b \to c \ell^- \bar{\nu}$ 
and $b \to u \ell^- \bar{\nu}$ decay widths, 
\begin{equation}
|V_{cb}| = 0.0411 \sqrt{
\frac{{\rm BR}(b \to c \ell^- \bar{\nu})}{0.105}} \,
\sqrt{ \frac{1.55~{\rm ps}}{\tau_b}}
~~~ \mbox{and} ~~~
|V_{ub}| = 0.00445 \sqrt{
\frac{{\rm BR}(b \to u \ell^- \bar{\nu})}{0.002}} \,
\sqrt{ \frac{1.55~{\rm ps}}{\tau_b}} \,,
\label{eq:vcb_vub}
\end{equation}
with total uncertainties estimated to be $\sim 5\%$.\,\cite{HF_preprint}
While the measurements of the inclusive $b \to \ell^-$ branching ratio 
and $b$-hadron lifetime are well established since 
several years, 
with current averages\,\cite{HF_preprint} of 
${\rm BR}(b \to \ell^-) = (10.58 \pm 0.07 \pm 0.17)\%$ and
$\tau_b = 1.564 \pm 0.014$~ps, 
analyses measuring the $b \to u \ell^- \bar{\nu}$ branching ratio 
are quite recent and unique to LEP. They face the difficulty of dealing with 
a very large $b \to c \ell^- \bar{\nu}$ background, but have the advantage 
to be sensitive to the whole lepton spectrum (rather than only to the 
its end-point where the $b \to c \ell^- \bar{\nu}$ decays are supressed).
L3, ALEPH, and DELPHI have now all published 
evidence for $b \to u\ell^-\bar{\nu}$ transitions, and their measurements 
average to\,\cite{HF_preprint} ${\rm BR}(b \to u\ell^-\bar{\nu}) =
(1.67 \pm 0.36 \pm 0.37 \pm 0.20)\times 10^{-3}$, 
where the first uncertainty summarizes statistics and experimental
systematics, the second uncertainty reflects the limited knowledge of
$b \to c\ell^-\bar{\nu}$ transitions, 
and the third one results from the modelling of $b \to u\ell^-\bar{\nu}$.
Using Eq.~\ref{eq:vcb_vub},
the LEP averages from inclusive semileptonic $b$ decays are
$|V_{cb}| = (40.8 \pm 0.4 _{\rm exp} 
\pm 2.0 _{\rm theory}) \times 10^{-3}$
and $|V_{ub}| = (4.04 ^{+0.62}_{-0.74}) \times 10^{-3}$. 
The former can be combined\,\cite{LEPHF}
with the less precise but consistent LEP estimate 
from exclusive decays to yield $|V_{cb}| = (40.5 \pm 1.8)\times 10^{-3}$.

\section{Results on the \boldmath $B_s^0$ decay width difference}
\label{sec:dgs}

\begin{figure}
\begin{minipage}{0.3825\textwidth}
\begin{center}
\epsfig{figure=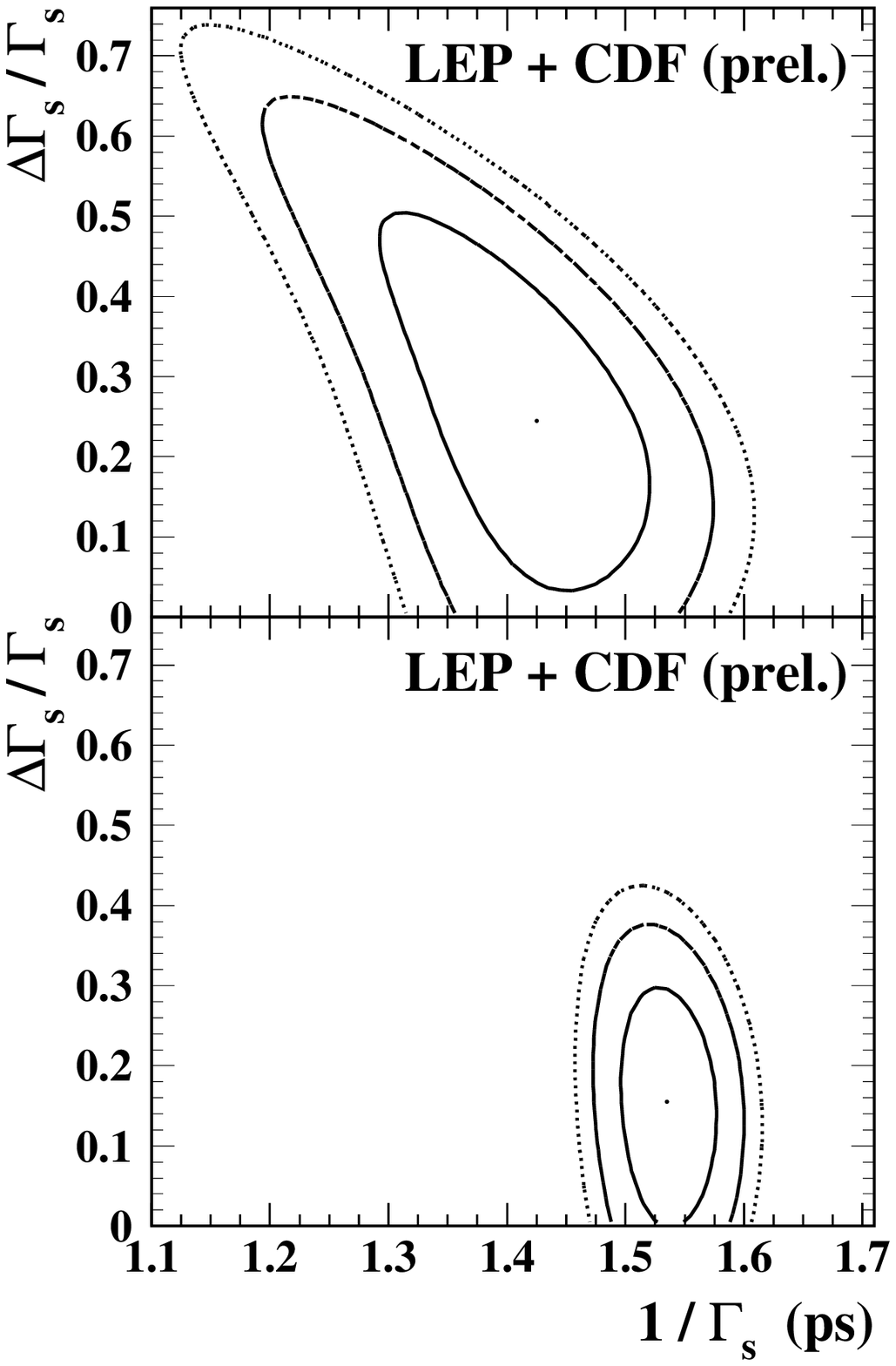,width=1.0\textwidth}
\caption{Combined 68\%, 95\% and 99\% CL contours in the plane 
($1/\Gamma_s$,$\Delta\Gamma_s/\Gamma_s$).\,\protect\cite{HF_preprint} 
The top (bottom) plot is obtained
without (with) the constraint $1/\Delta \Gamma_s = \tau_{B^0}$.} 
\label{fig:dgs}
\end{center}
\end{minipage}
\hfill 
\begin{minipage}{0.59\textwidth}
\begin{center}
\epsfig{figure=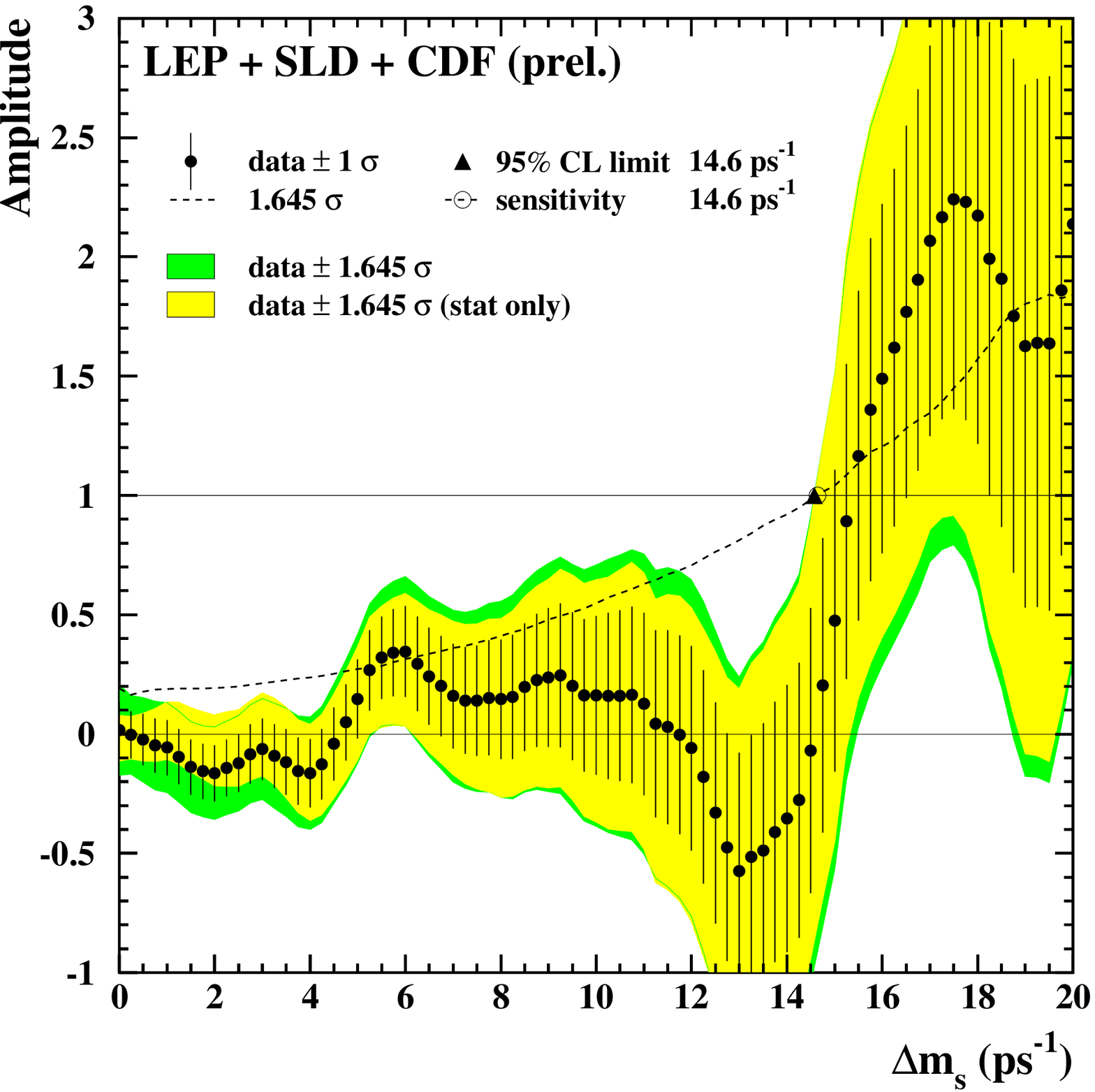,width=1.08\textwidth}
\caption{
Combined measurements of the $B_s^0$ oscillation amplitude as a 
function of $\Delta m_s$, 
obtained by the LEP $B$ oscillations working group.\,\protect\cite{LEPHF}
The measurements are dominated by statistical uncertainties. 
Neighbouring points are statistically correlated.}
\label{fig:dms}
\end{center}
\end{minipage}
\end{figure}

Information on $\Delta\Gamma_s$, the decay width difference between the 
two mass eigenstates of the $B_s^0\mbox{--}\bar{B}{_s^0}$ system, 
can be obtained by studying the proper time 
distribution of untagged data samples enriched in $B_s^0$ mesons.
In the case of an inclusive or a semileptonic $B_s^0$ decay selection, 
both the short- and long-lived components are present, 
and the proper time distribution is a superposition 
of two exponentials with decay constants 
$\Gamma_s\pm \Delta\Gamma_s/2$.
In principle, this provides sensitivity to both $\Gamma_s$ and 
$(\Delta\Gamma_s/\Gamma_s)^2$. Ignoring $\Delta\Gamma_s$ and fitting for 
a single exponential leads to an estimate of $\Gamma_s$ with a 
relative bias proportional to $(\Delta\Gamma_s/\Gamma_s)^2$. 
An alternative approach, which is directly sensitive to first order in 
$\Delta\Gamma_s/\Gamma_s$, 
is to determine the lifetime of $B_s^0$ candidates decaying to CP 
eigenstates; measurements exist for 
$B{^0_s} \to J/\psi \phi$,\,\cite{CDF_Jpsiphi} and now also for 
$B{^0_s} \to D_s^{(*)+} D_s^{(*)-}$,\,\cite{ALEPH-DGs} which are 
predicted to be mostly CP-even states.
Recently, ALEPH has also obtained for the first time an estimate of 
$\Delta\Gamma_s/\Gamma_s$ directly from a measurement of the 
$B{^0_s} \to D_s^{(*)+} D_s^{(*)-}$ branching ratio,\,\cite{ALEPH-DGs} 
under the assumption that 
these decays practically account for all the CP-even final states. 

Figure~\ref{fig:dgs} shows confidence contours 
in the plane $(1/\Gamma_s, \Delta\Gamma_s/\Gamma_s)$ obtained from
a combined likelihood built with all the available information from 
LEP and CDF, including 
dedicated $\Delta\Gamma_s$ studies as well as $B_s^0$ lifetime measurements. 
The corresponding results for $\Delta\Gamma_s/\Gamma_s$ are\,\cite{HF_preprint}
\begin{equation}
\Delta\Gamma_s/\Gamma_s = 0.24 ^{+0.16}_{-0.12} ~~~ \mbox{or} ~~~
\Delta\Gamma_s/\Gamma_s < 0.53 ~~~ \hbox{at 95\%~CL} 
\end{equation}
without external constraint, and
\begin{equation}
\Delta\Gamma_s/\Gamma_s = 0.17 ^{+0.09}_{-0.10} ~~~ \mbox{or} ~~~
\Delta\Gamma_s/\Gamma_s < 0.31 ~~~ \hbox{at 95\%~CL}  
\end{equation}
when constraining $1/\Gamma_s$ to the current world 
average of the $B^0$ lifetime. Such a constraint is well motivated 
theoretically, since the $B^0$ and $B_s^0$ decay widths are predicted to 
differ by $\sim1\%$ at most, 
but the current experimental check of this assumption,
$\tau_{B_s^0}/\tau_{B^0}=0.937 \pm 0.040$,\,\cite{HF_preprint} 
is still of limited precision.
These combined results on $\Delta\Gamma_s/\Gamma_s$ are not yet precise enough 
to test the Standard Model predictions, which typically lie 
between 5\% and 20\%.

\section{Search for \boldmath $B_s^0$ oscillations}
\label{sec:dms}

$B_s^0\hbox{--}\bar{B}{_s^0}$ oscillations have been the subject of many 
studies from ALEPH, DELPHI and OPAL, as well as SLD and CDF.
No oscillation signal has been found so far.
Because of the limited statistics available, 
the most sensitive analyses are currently the ones based on 
inclusive lepton samples, and on samples where a lepton and a $D_s$ meson 
have been reconstructed in the same jet. 
However, with larger samples, the most promising approach would be to use 
fully reconstructed $B_s^0$ mesons, which have a much better proper 
time resolution suitable to resolve higher oscillation frequencies.

DELPHI\,\cite{DELPHI-DGs-Dms} have fully reconstructed 44 $B_s^0$ candidates 
in the $\bar{D}^0 K^- \pi^+$, $\bar{D}^0 K^- a_1^+$, 
$D_s^{(*)-}\pi^+$  and $D_s^{(*)-}a_1^+$, 
channels, whereas ALEPH have recently reported 
50 candidates in the latter two channels. 
The number of signal events is estimated to be $\sim 20$ in each experiment, 
but with a proper time resolution of $\sim 0.08$~ps, more than 
two times better compared to more inclusive selections. 
As a result, these analyses, which have very poor sensitivity 
by themselves due to the lack of statistics, do nonetheless 
have a non-negligible impact on the average measurement of the oscillation 
amplitude ${\cal A}$ at high values of $\Delta m_s$, the mass difference 
between the two mass eigenstates of the $B_s^0$ system.

All results have been combined, including the latest ones from 
ALEPH\,\cite{ALEPH-Dms} released for this conference
and based on $D_s^- \ell^+$ correlations and 
fully reconstructed $B_s^0$ candidates, 
to yield the amplitudes ${\cal A}$ shown in Fig.~\ref{fig:dms}
as a function of $\Delta m_s$. 
In the combination procedure,\,\cite{HF_preprint}
the sensitivities of the inclusive lepton analyses, 
which depend directly 
on the assumed fraction $f_{B_s^0}$ of $B_s^0$ mesons in an unbiased sample 
of weakly-decaying $b$~hadrons, 
have been rescaled to a common value of 
$f_{B_s^0} = 0.096 \pm 0.012$.
This value is obtained from direct production measurements, 
measurements of the time-integrated mixing probability $\bar{\chi}$ 
of $b$-hadrons at LEP, as well as the new world average of the 
$B^0$ oscillation frequency, $\Delta m_d = 0.484 \pm 0.015$~ps.
The combined sensitivity for 95\%~CL exclusion of $\Delta m_s$ values 
is found to be 14.6~ps$^{-1}$, which is also the actual limit below 
which all values of $\Delta m_s$ are excluded by the data at 95\%~CL.
No oscillation signal can be claimed based on the 
deviation from ${\cal A}=0$ seen in Fig.~\ref{fig:dms} around 17~ps$^{-1}$:
a fast Monte Carlo study\,\cite{Abbaneo_Boix} shows indeed that statistical 
fluctuations can produced a more significant deviation (anywhere in the 
explored range in $\Delta m_s$) in $\sim 3\%$ of the 
samples generated with a very large true value of $\Delta m_s$.

\begin{figure}
\vspace{-0.5cm}
\begin{center}
\epsfig{figure=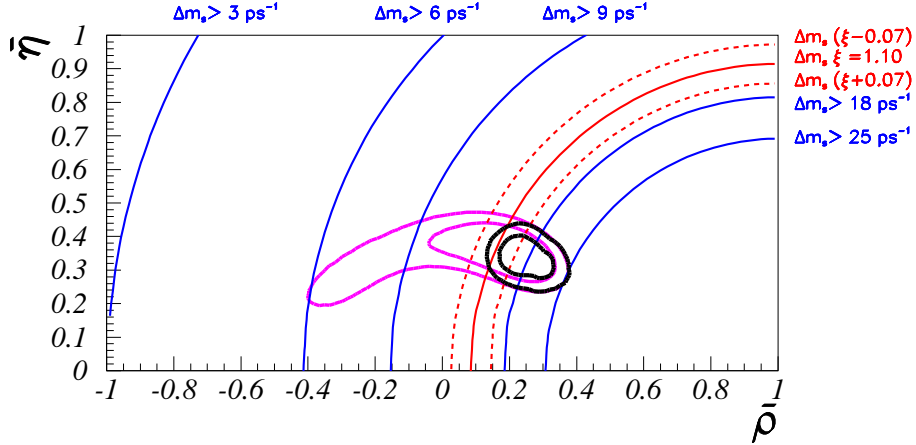,width=0.75\textwidth}
\end{center}
\vspace{-0.5cm}
\caption{
Results of two different CKM fits
shown in the $(\bar{\rho},\bar{\eta})$ plane, 
using the Wolfenstein parametrization.
The two dark closed curves represent the 68\% and 95\% CL contours
obtained using the procedure and assumptions 
of Caravaglios et al.\,\protect\cite{CKM_fits} 
with $\Delta m_d = 0.484 \pm 0.015$~ps and the $\Delta m_s$ results 
of Fig.~\protect\ref{fig:dms}.
The two lightest closed curves are the result of the same fit,
except that the experimental information on $\Delta m_s$ is ignored.
The circles centered on (1,0) indicate the constraints 
corresponding to various values of $\Delta m_s$, with the hadronic 
uncertainty indicated as the dashed band.}
\label{fig:CKM_fits}
\end{figure}
The information on $|V_{ts}|$ obtained, 
in the framework of the Standard Model, from the combined $\Delta m_s$ 
limit is hampered by the hadronic uncertainty, as is the case when extracting 
$|V_{td}|$ from $\Delta m_d$.
However, many uncertainties cancel in the frequency ratio
\begin{equation}
\frac{\Delta m_s}{\Delta m_d} = \frac{m_{B_s^0}}{m_{B^0}}\, \xi^2
                    \left|\frac{V_{ts}}{V_{td}}\right|^2 \,,
\label{eq:ratio} 
\end{equation}
where $\xi$ is currently known to $\sim 6\%$ from lattice QCD.
This relation can be used in fits of the CKM matrix,
together with the experimental results on 
$\Delta m_s$,
$\Delta m_d$,
$|V_{ub}/V_{cb}|$ and $\epsilon_K$,
as well as theoretical inputs and unitarity constraints.
Examples of such fits,\,\cite{CKM_fits} shown 
in Fig.~\ref{fig:CKM_fits}, illustrate the fact that the combined 
$\Delta m_s$ results from Fig.~\ref{fig:dms} provide, 
together with the measured value of $\Delta m_d$, a significant 
constraint on the CKM matrix, favouring positive values 
of the Wolfenstein parameter $\rho$.

\section{CP violation in \boldmath $B^0 \to J/\psi K^0_S$ decays}
\label{sec:cpv}

ALEPH has recently released a new measurement\,\cite{ALEPH-sin2beta} of the 
CP asymmetry in $B^0,\bar{B}{^0} \to J/\psi K^0_S$ decays,  
\begin{equation}
A(t) = \frac{N_{B^0}(t) - N_{\bar{B}{^0}}(t)} 
            {N_{B^0}(t) + N_{\bar{B}{^0}}(t)} 
     = - \sin(2\beta)\sin(\Delta m_d \, t) \,,
\end{equation}
where $N_{B^0}(t)$ and $N_{\bar{B}{^0}}(t)$
are the number of events produced as $B^0$ and $\bar{B}{^0}$ 
as a function of the proper time $t$, and $\beta$ is one of the angles 
of the CKM unitarity triangle. From a sample of
23 fully reconstructed candidates, selected with a signal efficiency 
of $(28\pm2)\%$ and an estimated purity of $(71\pm 12)\%$, ALEPH measures
$\sin(2\beta)=
0.93{^{+0.64}_{-0.88}}\mbox{(stat)}{^{+0.36}_{-0.24}}\mbox{(syst)}$.
The systematic uncertainty is dominated by the 
limited knowledge of the probability of mistagging the initial state, 
measured to be $(25\pm 6)\%$ using $B^\pm \to J/\psi K^\pm$ events.
This $\sin(2\beta)$ result can be combined with previous 
measurements from OPAL and CDF to yield\,\cite{ALEPH-sin2beta}
$\sin(2\beta)=0.91 \pm 0.35$ or $\sin(2\beta)>0$ at 98.5\% CL, 
increasing the confidence that CP violation has been observed in the 
$B$ sector.

\section*{Acknowledgments}
I would like to thank D.~Abbaneo and 
R.~Forty for useful comments and careful reading 
of this writeup, A.~Stocchi for providing Fig.~\ref{fig:CKM_fits}, and
the organizers of these ``Rencontres'' for an enjoyable conference.

\newcommand{\auth}[1]{#1,}
\newcommand{\coll}[1]{#1 coll.,}
\newcommand{\authcoll}[2]{#1 \etal, \coll{#2}}
\newcommand{\init}[1]{#1.\,}
\newcommand{\etal}{et al.}
\newcommand{\titl}[1]{``#1'',}
\renewcommand{\titl}[1]{$\!\!$}
\newcommand{\J}[4]{{#1}{\bf #2}, #3 (#4)}
\newcommand{\subJ}[1]{subm.\,to #1}
\newcommand{\PRL}[3]{\J{Phys.\,Rev.\,Lett.\ }{#1}{#2}{#3}}
\newcommand{\subPRL}{\subJ{Phys.\,Rev.\,Lett}}
\newcommand{\PRD}[3]{\J{Phys.\,Rev.\,D}{#1}{#2}{#3}}
\newcommand{\subPRD}{\subJ{Phys.\,Rev.\,D}}
\newcommand{\ZPC}[3]{\J{Z.\,Phys.\,C}{#1}{#2}{#3}}
\newcommand{\subZPC}{\subJ{Z.\,Phys.\,C}}
\newcommand{\PLB}[3]{\J{Phys.\,Lett.\,B}{#1}{#2}{#3}}
\newcommand{\EPJC}[3]{\J{Eur.\,Phys.\,J.\,C}{#1}{#2}{#3}}
\newcommand{\subPLB}{\subJ{Phys.\,Lett.\,B}}
\newcommand{\NPB}[3]{\J{Nucl.\,Phys.\,B}{#1}{#2}{#3}}
\newcommand{\subNPB}{\subJ{Nucl.\,Phys.\,B}}
\newcommand{\NIM}[3]{\J{Nucl.\,Instrum.\,Methods A}{#1}{#2}{#3}}
\newcommand{\subNIM}{\subJ{Nucl.\,Instrum.\,Methods A}}
\newcommand{\JHEP}[3]{\J{J.\,of High Energy Physics }{#1}{#2}{#3}}
\newcommand{\ARNPS}[3]{\J{Ann.\,Rev.\,Nucl.\,Part.\,Sci.\ }{#1}{#2}{#3}}
\newcommand{\newref}{\\}


\section*{References}
\begin{thebibliography}{99}

\bibitem{LEPHF}
The LEP $|V_{cb}|$, $|V_{ub}|$, $b$-lifetimes,
$\Delta\Gamma_s$ and $B$ oscillations working groups
are coordinated by the LEP Heavy Flavour Steering Group; 
see \verb+http://www.cern.ch/LEPHFS/+ and the links therein.
The combination procedures used by these groups are 
described in a public document.\,\cite{HF_preprint}

\bibitem{OPAL-search}
\authcoll{\init{G}Abbiendi}{OPAL}
\titl{Search for new physics in rare $B$ decays}
\PLB{476}{233}{2000}.

\bibitem{LEPHFWG_1998}
\auth{The LEP heavy flavour working group} 
\titl{Input parameters for the LEP electroweak heavy flavour results 
for summer 1998 conferences}
LEPHF 98-01, Sept.\ 1998.

\bibitem{ALEPH-bfrag}
\coll{ALEPH}
\titl{Measurement of the effective $b$ quark fragmentation 
function at the $Z$ peak} 
ALEPH 2000-020 CONF 2000-017, March 2000.

\bibitem{SLD-bfrag}
\authcoll{\init{K}Abe}{SLD}
\titl{Precise measurement of the $b$ quark fragmentation function 
in $Z^0$ boson decays}
SLAC-PUB-8316, hep-ex/9912058, Dec.\ 1999, \subPRL.

\bibitem{Vcb-param}
\auth{\init{I}Caprini, \init{L}Lellouch and \init{M}Neubert}
\titl{Dispersive bounds on the shape of 
$\bar{B} \to D^{(*)}\ell \bar{\nu}$ form factors}
\NPB{530}{153}{1998}.

\bibitem{Vcb-OPAL-new}
\authcoll{\init{G}Abbiendi}{OPAL}
\titl{Measurement of $|V_{cb}|$ using $\bar{B}{^0} \to D^{*+} \ell^- 
\bar{\nu}$ decays} 
CERN-EP/2000-032, Feb. 2000, 
\subPLB.

\bibitem{HF_preprint}
\authcoll{ALEPH, CDF, DELPHI, L3, OPAL}{SLD}
\titl{Combined results on $b$-hadron
production rates, lifetimes, oscillations, and semileptonic decays}
LEPHFS note 99-02, April 2000, to be subm.\ as CERN-EP preprint, 
and references therein.

\bibitem{CDF_Jpsiphi}
\authcoll{\init{F}Abe}{CDF}
\titl{Measurement of $b$ hadron lifetimes using $J/\psi$ final states at CDF}
\PRD{57}{5382}{1998}.

\bibitem{ALEPH-DGs}
\authcoll{\init{R}Barate}{ALEPH}
\titl{A study of the decay width difference in the 
$B^0_s\mbox{--}\bar{B}{^0_s}$ system using $\phi \phi$ correlations} 
CERN-EP/2000-036, Feb.\ 2000, \subPLB.

\bibitem{DELPHI-DGs-Dms}
\coll{DELPHI}
\titl{Search for $B_s^0\mbox{--}\bar{B}{_s^0}$ oscillations, 
measurement of the $B_s^0$ lifetime and limit on the lifetime difference}
DELPHI 99-109 CONF 296, June 1999,
contribution 4\_520 to Int.\ Europhysics Conf.\ 
on High Energy Physics, Tampere, 1999. 

\bibitem{ALEPH-Dms}
\coll{ALEPH}
\titl{Study of $B^0_s\mbox{--}\bar{B}{^0_s}$ oscillations using fully 
reconstructed $B^0_s$ and $D^-_s \ell^+$ events} 
ALEPH 2000-029 CONF 2000-024, March 2000.

\bibitem{Abbaneo_Boix}
\auth{\init{D}Abbaneo and \init{G}Boix} 
\JHEP{8}{4}{1999}.

\bibitem{CKM_fits}
\auth{\init{F}Caravaglios \etal}
hep-ph/0002171, 
presented by \init{A}Stocchi at the 3rd Int.\ Conf.\ on 
$B$ physics and CP violation, Taipei, Taiwan, Dec.\ 1999.

\bibitem{ALEPH-sin2beta}
\coll{ALEPH}
\titl{Study of the CP asymmetry of $B^0 \to J/\psi K^0_S$ decays in ALEPH}
ALEPH 99-099 CONF 99-054, Nov.\ 1999, and references therein.

\end{thebibliography}
\end{document}